\documentclass[aps,prl,reprint,footinbib]{revtex4-1}
\usepackage{blindtext}
\usepackage{graphicx}
\usepackage{epstopdf}

\newcommand{\beq}{\begin{equation}}
\newcommand{\eeq}{\end{equation}}
\newcommand{\bea}{\begin{eqnarray}}
\newcommand{\eea}{\end{eqnarray}}

\newcommand{\hi}{H{\sc i}~}
\newcommand{\HI}{H{\sc i}}

\usepackage[colorlinks,urlcolor=blue,citecolor=blue,linkcolor=blue]{hyperref}
\usepackage{gensymb}
\usepackage{amsmath}
\usepackage{color}
\usepackage[titletoc,title]{appendix}
\usepackage{cleveref}


\begin{document}
\title{Neutral hydrogen structures trace dust polarization angle: Implications for cosmic microwave background foregrounds}

\author{Clark, S.E.$^{1}$}
\author{Hill, J. Colin$^{1}$}
\author{Peek, J.E.G.$^{2}$}
\author{Putman, M.E.$^{1}$}
\author{Babler, B.L.$^{3}$}

\affiliation{$^{1}$Department of Astronomy, Columbia University, New York, NY}
\affiliation{$^{2}$Space Telescope Science Institute, Baltimore, MD}
\affiliation{$^{3}$University of Wisconsin at Madison, Madison, WI}

\begin{abstract}
Using high-resolution data from the Galactic Arecibo L-Band Feed Array HI (GALFA-\HI) survey, we show that linear structure in Galactic neutral hydrogen (\HI) correlates with the magnetic field orientation implied by \textit{Planck} 353 GHz polarized dust emission. The structure of the neutral interstellar medium is more tightly coupled to the magnetic field than previously known. At high Galactic latitudes, where the \textit{Planck} data are noise-dominated, the \hi data provide an independent constraint on the Galactic magnetic field orientation, and hence the local dust polarization angle. We detect strong cross-correlations between template maps constructed from estimates of dust intensity combined with either \HI-derived angles, starlight polarization angles, or \textit{Planck} 353 GHz angles. The \hi data thus provide a new tool in the search for inflationary gravitational wave $B$-mode polarization in the cosmic microwave background, which is currently limited by dust foreground contamination. 
\end{abstract} 
 
\maketitle

The cosmic microwave background (CMB) is the pervasive residual radiation from the formation of the Universe. The detection of primordial $B$-mode polarization in the CMB is a major goal of contemporary cosmology. 
This signal is imprinted at the surface of last scattering by perturbations from gravitational waves generated during the epoch of inflation, a period of rapid expansion in the early Universe \cite{1997PhRvL..78.2054S, 1997ApJ...482....6S, Kamionkowski:1997hq}. An inflationary gravitational wave (IGW) $B$-mode measurement would be the first direct evidence of inflation. A number of experiments are pursuing the signal, using ground-based (e.g. ABS \cite{EssingerHileman:2010uf}; Advanced ACT \cite{Niemack:2010fk}; BICEP2/\textit{Keck} Array \cite{Ade:2015wh}; CLASS \cite{EssingerHileman:2014wg}; POLARBEAR \cite{Kermish:2012wh}; SPT-3G \cite{Benson:2014br}), balloon-borne (EBEX \cite{ReichbornKjennerud:2010vb}; SPIDER \cite{Fraisse:2011jv}), and space telescopes (\textit{Planck} \cite{Collaboration:2015tu}). 

Unfortunately, our view of the polarized CMB is obscured by contaminating foregrounds. For IGW $B$-mode searches at frequencies $\gtrsim 100$ GHz, the largest foreground is Galactic polarized dust emission. Aspherical dust grains in the Milky Way align their short axes with the ambient magnetic field, and interstellar radiation is absorbed and reradiated by the dust as partially polarized light. The BICEP2 collaboration claimed a measurement of primordial $B$-modes \cite{Ade:2014dj}, but subsequent analyses determined that the detection could be attributed entirely to Galactic dust \cite{Flauger:2014ub,Collaborations:2015wh}. A detailed understanding of the foreground polarization signal is required before a definitive IGW $B$-mode detection can be achieved. Pursuant to that goal, the \textit{Planck} satellite recently mapped the full sky at 353 GHz, a frequency dominated by thermal dust emission. These data can be used to subtract the foreground polarization pattern from lower-frequency CMB observations. To optimize the chance of primordial $B$-mode detection, experiments should target the ``cleanest" regions of sky: areas where there is relatively little polarized dust, and where the dust polarization structure is measured with high signal-to-noise. The \textit{Planck} maps are limited in this regard, because the \textit{Planck} polarized signal is noise-limited at high Galactic latitudes, where the dust column is lowest. Thus IGW $B$-mode searches are plagued by a trade-off: the regions of lowest foreground amplitude are also the regions with the poorest foreground constraints.

We present an entirely new method for constraining Galactic foregrounds. Using only the morphology of diffuse neutral hydrogen (\HI) structures, we predict the orientation of polarized dust emission at high precision over a range of angular scales. In parallel with existing measurements of polarized CMB foregrounds, our recovery of the dust polarization angle will increase the precision of foreground models. This is especially valuable in regions where the \textit{Planck} 353 GHz data are noise-limited.

\begin{figure*}
\centering
\includegraphics[width=\textwidth]{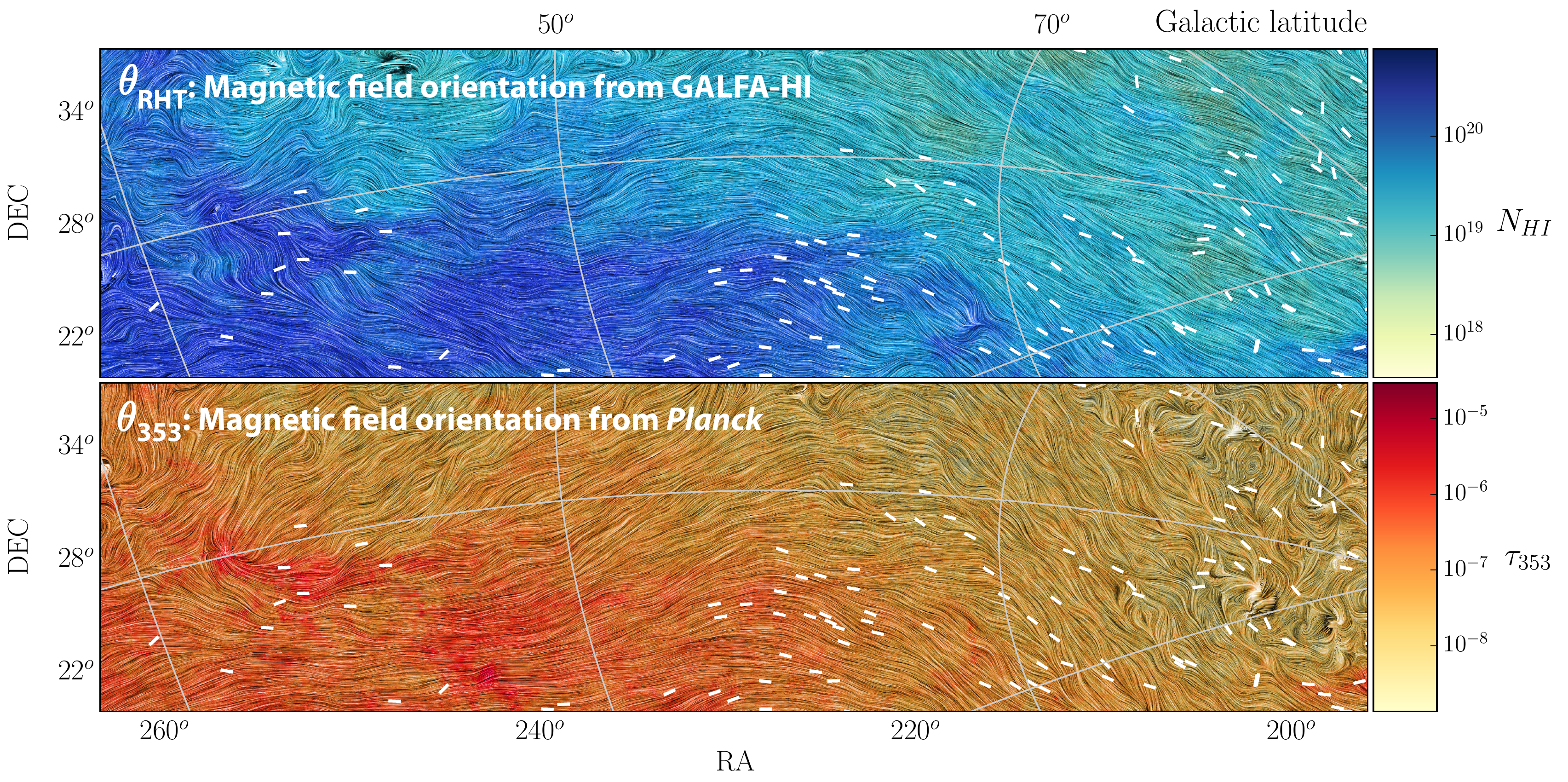}
\caption{Plane-of-sky magnetic field orientation as predicted by $\theta_{RHT}$ (top) and $\theta_{353}$ (bottom). Color maps are integrated HI column density from $v = -61.5$ km s$^{-1}$ to $+61.5$ km s$^{-1}$ ($N_{HI} \,[{\rm cm}^{-2}]$), and dust opacity ($\tau_{353}$). \textit{Planck} and RHT $Q$ and $U$ values are smoothed with a FWHM = $1\degree$ Gaussian kernel, then used to construct $\theta_{353}$ and $\theta_{RHT}$, which are visualized using line integral convolution (LIC; \cite{Cabral:1993}). The high latitude ($b \gtrsim 70$) behavior of the $\theta_{353}$ LIC pattern is due to \textit{Planck} noise. White pseudovectors represent starlight polarization angles. Galactic latitude lines lie at $b = 30\degree$, $50\degree$, $70\degree$, from left to right. Galactic longitude lines lie at $l = 80\degree$, $50\degree$, $20\degree$, from top to bottom. \label{colormaps}}
\end{figure*}

This work follows the discovery that linear structures in \hi are elongated in the direction of the interstellar magnetic field as probed by starlight polarization \cite{Clark:2014it}. Here, we demonstrate that \hi orientation is well correlated with the \textit{Planck} 353 GHz polarization angle across a region of high Galactic latitude sky. 
Note that the \textit{Planck} data enable quantitative conclusions beyond the previous work, which considered only 153 sparsely sampled starlight polarization measures over 1,300 deg$^2$ of sky. Also, polarized dust emission samples the full line of sight, whereas starlight polarization only traces the magnetic field out to the distance of the star. The relationship between dust and \hi in the interstellar medium (ISM) is deeper than their correlation in column density (e.g. \cite{Burstein:1982wb}), which is already used to estimate the amplitude of polarized dust emission \cite{Flauger:2014ub, Collaboration:2014wt}. Small dust grains and long plumes of \hi are both aligned by the magnetic field, though the mechanism for aligned \hi structure formation is not yet well understood.

The slender linear features that best trace the orientation of the Galactic magnetic field are only revealed by high spatial and spectral resolution \hi maps (see \cite{Clark:2014it} for details). We use data from the Galactic Arecibo L-Band Feed Array \hi survey (GALFA-\HI; \cite{Peek:2011fp}) with the Arecibo 305m radio antenna. GALFA-\hi has an angular resolution of FWHM $\simeq 4'$, a spectral resolution of $0.18$ km s$^{-1}$, and a brightness temperature noise of $\sim 140$ mK rms per 1 km s$^{-1}$ integrated channel over 13,000 deg$^2$ of sky. This work uses data from the forthcoming second data release \cite{Peek:2015}.

We analyze 353 GHz polarization data obtained by the \textit{Planck} satellite's High Frequency Instrument (HFI) \cite{Collaboration:2014wt}. These data have an angular resolution of FWHM $\simeq 5'$, comparable to GALFA-\HI. We transform the \textit{Planck} data from Galactic to Equatorial coordinates \cite{Healpix:footnote}. For all analyses, we apply a mask constructed from the union of all point source masks provided for each HFI channel in both temperature and polarization.

We additionally consider 126 optical starlight polarization measures in this region \cite{Heiles:2000un}. Starlight is polarized parallel to the magnetic field by the preferential absorption of aligned grains.

We quantify the orientation of GALFA-\hi structures using the Rolling Hough Transform (RHT), a machine vision technique \cite{Clark:2014it}. The RHT runs on image data, and outputs $R\left(\theta\right)$, linear intensity as a function of angle, for every pixel in the input map. For a detailed description of the RHT we refer the reader to \cite{Clark:2014it}.

For this work we select a 1,278 deg$^2$ region of the GALFA-\hi sky. The region, which spans right ascension 195$\degree$ to 265$\degree$ and declination 19.1$\degree$ to 38.3$\degree$, stretches from $b = 30\degree$ above the Galactic plane to $b =  81.7\degree$, nearly Galactic zenith. We analyze this GALFA-\hi region from $-13.5$ km s$^{-1}$ to $+13.5$ km s$^{-1}$, binned in $3.0$ km s$^{-1}$ integrated velocity channels. 

Linear polarization data can be fully described by either a polarization angle $\psi$ and polarized intensity $P$ or by the Stokes parameters $Q$ and $U$, where $\psi = 1/2 \, \mathrm{arctan} (U/Q)$ and $P^2 = Q^2 + U^2$. We define from the RHT output 

\bea
Q_{RHT} & = & \int \mathrm{cos} \left(2 \theta \right) \cdot R\left(\theta \right) \, d\theta \,  \nonumber \\
U_{RHT} & = & \int \mathrm{sin} \left(2 \theta \right) \cdot R\left(\theta \right) \, d\theta, \,
\eea
where values are calculated for each point in the image data. We process each velocity channel with the RHT and add the resulting $Q_{RHT}$ and $U_{RHT}$ maps. 

We define $\theta_{RHT} = 1/2 \, \mathrm{arctan} (U_{RHT}/Q_{RHT})$, an estimate for the orientation of the magnetic field derived solely from \hi data. We compare this value to $\theta_{353}$, a $90\degree$ rotation of the polarization angle obtained from $Q_{353}$ and $U_{353}$ (we use the IAU polarization definition). The polarization angle of dust emission is conventionally taken to be $90\degree$ from the orientation of the local Galactic magnetic field (however, see \cite{Lazarian:2007do} and references therein).

\begin{figure}
\centering
\includegraphics[width=0.5\textwidth]{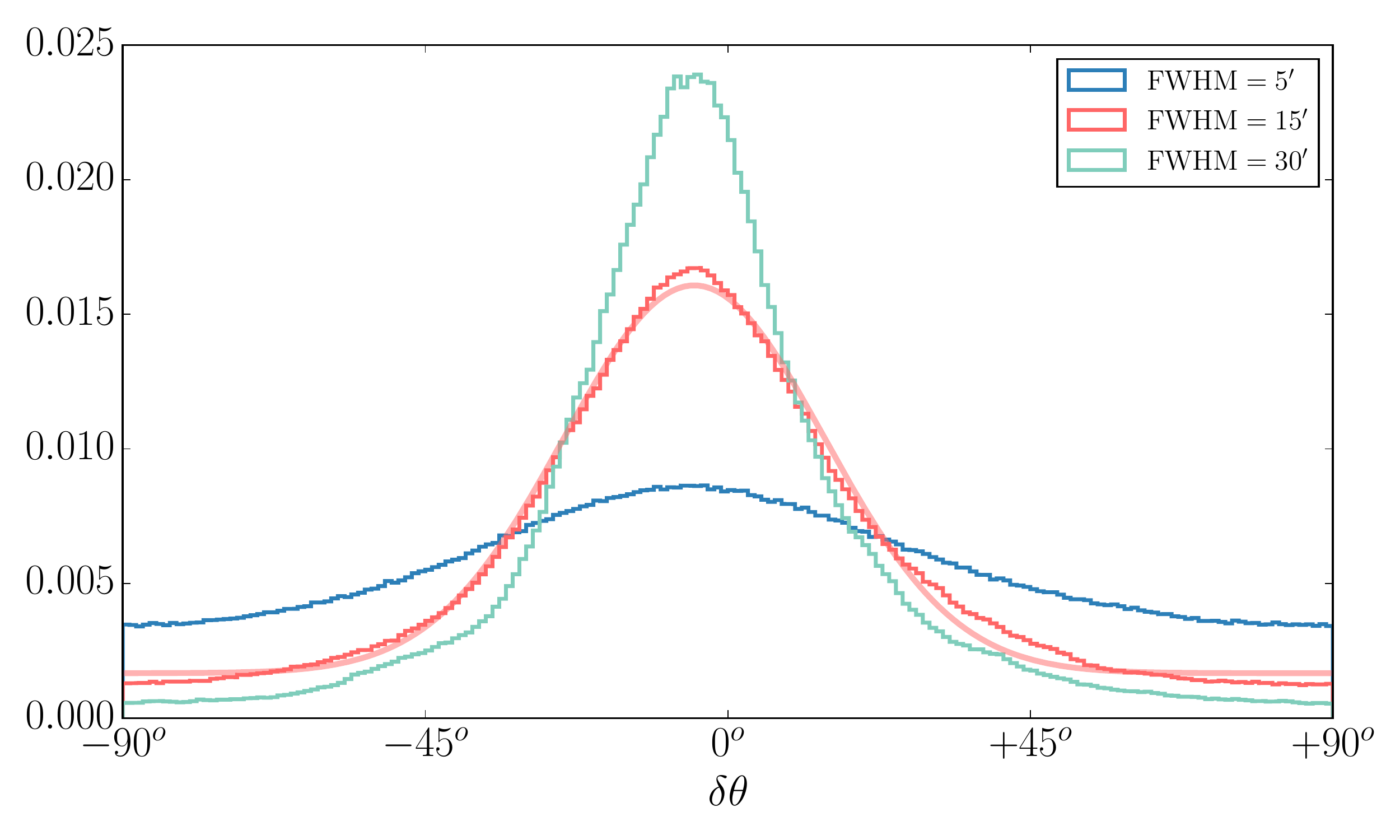}
\caption{Normalized histograms of $\delta \theta = \theta_{353} - \theta_{RHT}$ in $1\degree$ bins at resolutions of FWHM = $5'$, $15'$, and $30'$. The Gaussian fit to the FWHM = $15'$ histogram shown has a standard deviation $\sigma = 19.4\degree$. 
\label{smoothhists}}
\end{figure}

We calculate $\theta_{353}$ and $\theta_{RHT}$ for the region described. Figure \ref{colormaps} shows a map of each of these quantities on the sky, along with starlight polarization angles. Although derived from independent data, these three estimates for the plane-of-sky magnetic field orientation trace one another remarkably well. Figure \ref{smoothhists} shows histograms of $\delta \theta \equiv \theta_{353} - \theta_{RHT}$. We construct $\theta_{353}$ and $\theta_{RHT}$ from $Q$ and $U$ maps smoothed to three different resolutions. For Gaussian smoothing kernels of FWHM = $5'$, $15'$, and $30'$, we find Gaussian fits to the $\delta \theta$ histogram with standard deviation $\sigma = 30.2\degree$, $\sigma = 19.4\degree$, and $\sigma = 14.4\degree$, respectively. We run a Monte Carlo analysis to determine the pure \textit{Planck} noise contribution to $\delta \theta$, and find this noise is responsible for a Gaussian component with $\sigma = 16.0\degree$, $\sigma = 6.1\degree$, and $\sigma = 3.5\degree$ for each respective smoothing kernel. Thus as the data are smoothed to larger angular scales, $\theta_{353}$ and $\theta_{RHT}$ obtain ever better agreement, and a non-negligible fraction of the $\delta \theta$ scatter is solely due to $Q_{353}$ and $U_{353}$ measurement noise. The $\delta \theta$ histograms are centered at about $-3\degree$ to $-4\degree$. This small offset from zero may be due to either residual systematics in the $353$ GHz map \cite{Collaboration:2015ta} or true systematic differences between $\theta_{353}$ and $\theta_{RHT}$.

To further characterize the relationship between RHT, \textit{Planck}, and starlight polarization angles, we construct simple template maps and compute cross-power spectra between them. We construct the templates using the \textit{Planck} 353 GHz intensity, $I_{353}$. A full polarization template would also require an estimate of the polarization fraction, $p=P/I$, but since our goal is to isolate the polarization angle information, we set $p=1$ in all templates. (Over a small patch of sky, $p \approx$ constant is a reasonable approximation, and one can simply re-scale our power spectra for a given value of $p$.) Furthermore, measuring $P$ from the \textit{Planck} data is non-trivial, as simple estimators are noise-biased (e.g. \cite{Plaszczynski:2013gg}). The templates are 
\beq
Q = I_{353} \cos(2\psi) \,\,\, \mathrm{and} \,\,\, U = I_{353} \sin(2\psi),
\label{eq.QUtemplate}
\eeq
where $\psi$ is either the RHT, \textit{Planck}, or starlight polarization angle. For all templates, we smooth the $Q$ and $U$ data to a common resolution of FWHM = 4$\degree$ before computing $\psi$. This prohibits small-scale noise in $Q$ and $U$ from contaminating the templates on large scales via the harmonic-space convolution implied by the real-space map multiplication in Eq.~(\ref{eq.QUtemplate}). 
To avoid noise biases, we measure cross-correlations between templates constructed from independent half-mission splits of the \textit{Planck} data.

We apply a common mask to all template maps, consisting of the \textit{Planck} point source mask and a mask removing regions that are more than $7\degree$ from starlight data, regions where the integrated RHT intensity is zero, and the edges of the region, where RHT artifacts could arise. The total unmasked sky subtends 1,181 deg$^2$, 92\% of the original area. We apodize the mask with a Gaussian taper of FWHM = 15$'$. We use {\tt polspice} \cite{Chon:2004tp} to compute $EE$ and $BB$ power spectra ($C_{\ell}$, where multipole $\ell$ is the harmonic variable conjugate to angular scale), corresponding to the usual curl-free and divergence-free decompositions of polarization data \cite{1997PhRvL..78.2054S, Kamionkowski:1997hq}, respectively. We calibrate the {\tt polspice} internal parameters using 100 simulations of polarized dust power spectra with properties matching recent \textit{Planck} measurements \cite{Collaboration:2014wt, Collaboration:2014uh}. We bin the measured power spectra in four logarithmically spaced multipole bins between $\ell=40$ and $\ell=600$ (centered at $\ell=59, 116, 229$, and $451$).  Error bars are calculated in the Gaussian approximation from the auto-power spectra of the template maps used in each cross-correlation.  Sample variance is not included in the error bars, as our interest is in comparing measurements of the same modes on the sky.

\begin{figure}
\centering
\includegraphics[width=0.5\textwidth]{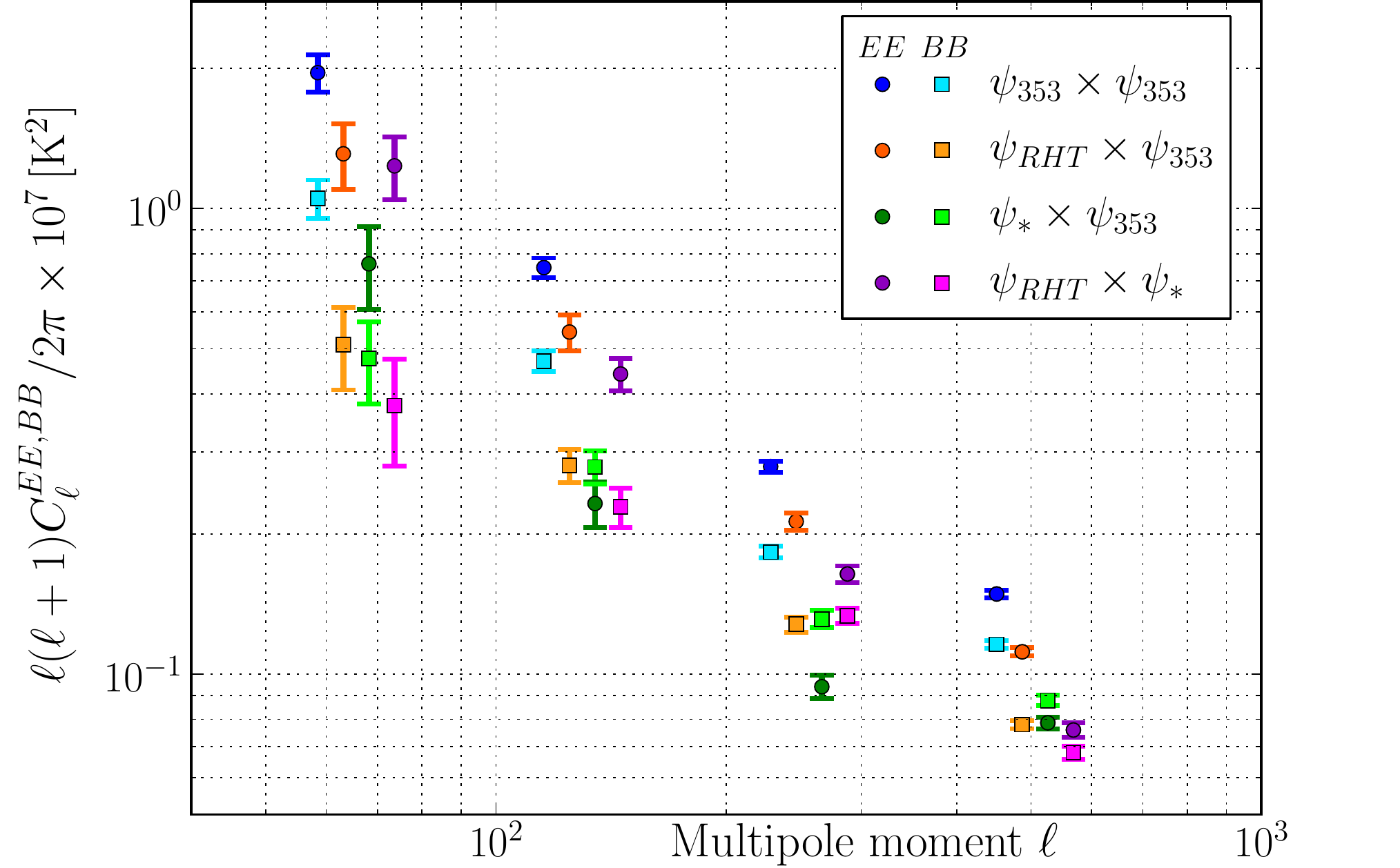}
\caption{Cross-power spectra of polarization template maps constructed from $I_{353}$ and either \textit{Planck} ($\psi_{353}$), RHT ($\psi_{RHT}$), or starlight polarization ($\psi_*$) data (Eq.~\ref{eq.QUtemplate}). Shown are $E$-mode (circles) and $B$-mode (squares) components. Significant ($40$--$70\sigma$) cross-correlations are detected in all cases. \label{cps}}
\end{figure}

Figure \ref{cps} shows cross-power spectra for the template maps constructed from RHT, \textit{Planck}, and starlight data.  We refrain from fitting a model to the data, as we have not considered $p$ in our templates, but instead consider the relative amplitudes of the cross-power spectra. 
For the \textit{Planck}-only templates and the \textit{Planck}--RHT, \textit{Planck}--starlight, and RHT--starlight cross-correlations, respectively, we detect the $E$-mode power spectrum at $70\sigma$, $55\sigma$, $40\sigma$, and $40\sigma$ significance. We detect the $B$-mode power spectrum at $65\sigma$, $60\sigma$, $50\sigma$, and $40\sigma$ significance. We verify that template maps constructed with random angles yield a cross-power spectrum consistent with zero (even when using the true $I_{353}$ data in the random-angle templates). We compare the template cross-power spectra with the actual $EE$ and $BB$ power spectra measured directly from $Q_{353}$ and $U_{353}$ and infer a mean $p \sim 5\%$, which is reasonable for this region \cite{Collaboration:2014wt}.

Although significant cross-correlations are detected for all templates in Figure~\ref{cps}, the \textit{Planck}-only templates yield higher amplitudes than the cross-correlations with RHT- or starlight-based templates. While this could be due to physical differences between angles, we note that the RHT--\textit{Planck} and RHT--starlight cross-correlations yield similar results (especially at low-$\ell$), suggesting that the \textit{Planck}-only templates' power spectra could be systematically biased. Because the angle construction relies on the $U_{353}/Q_{353}$ ratio, it is sensitive to any effect that modifies the zero point of the maps. Such effects could include gain calibration drifts or intensity-to-polarization leakage that varies over the sky, both of which are known to be present in the \textit{Planck} data at some level \cite{Collaboration:2015ta}. Indeed, scan-synchronous systematics have been detected in \textit{Planck} temperature data \cite{Collaborations:2015y,2013PhRvD..88j1301K}, and maps of $\delta \theta = \theta_{353}-\theta_{RHT}$ present clear visual evidence of residuals that are highly correlated with the \textit{Planck} scan directions. We leave a detailed consideration of these systematic effects on the \textit{Planck} angles to future work. Note that direct measurements of $EE$ and $BB$ power spectra from $Q_{353}$ and $U_{353}$ are more immune to these systematics than the angle construction, but we require the latter method to compare \textit{Planck} data in a straightforward way to the RHT- and starlight-based templates.

\begin{figure}
\centering
\includegraphics[width=0.5\textwidth]{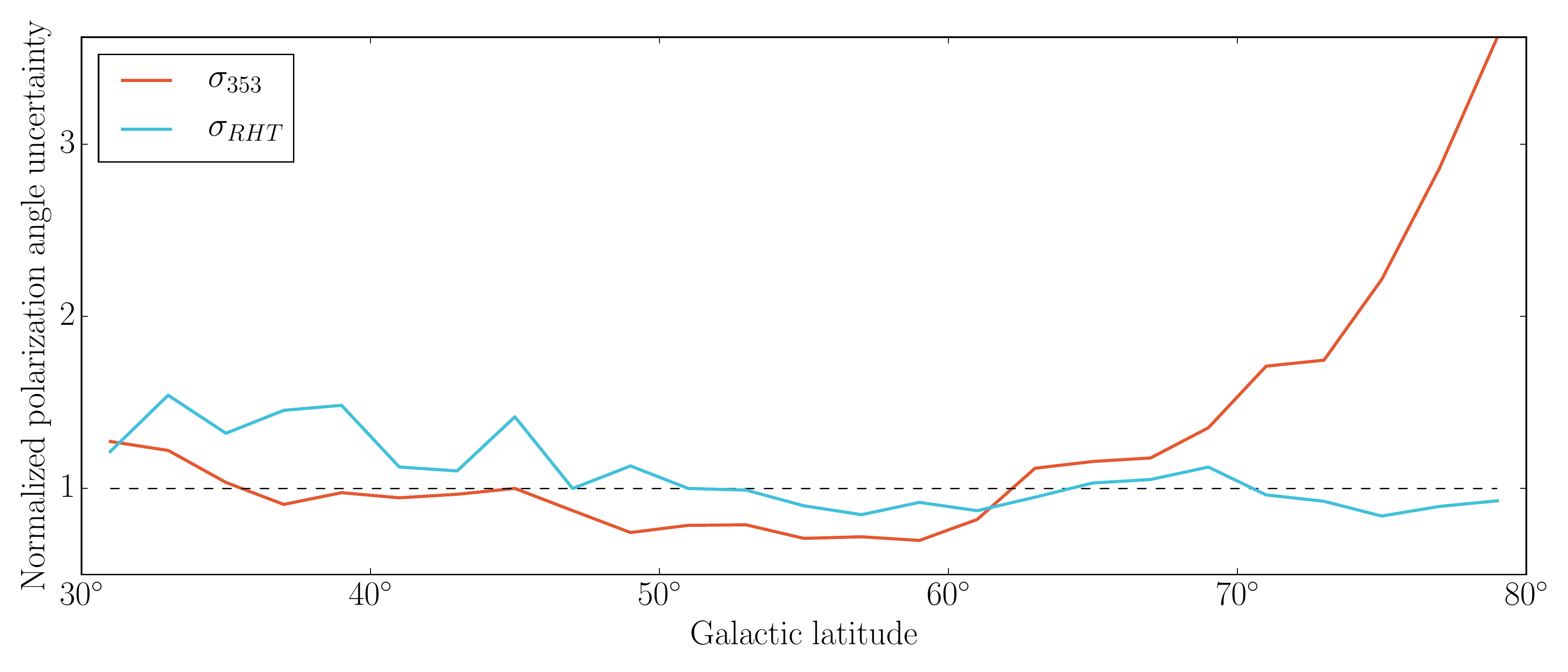}
\caption{Angle uncertainties averaged over $2\degree$ Galactic latitude bins, normalized by their respective median values (dashed line). \label{sigmas}}
\end{figure}

The RHT--\textit{Planck} cross-power spectra yield an amplitude ratio $C_l^{EE}/C_l^{BB} \approx 2$, a result consistent with the \textit{Planck} 353 GHz measurement \cite{Collaboration:2014uh}, though this must be interpreted with caution as we have not modeled $p$ in our templates. Many current models of the dust polarized sky \cite{ODea:2011bm, Delabrouille:2013wf} predict equal $E$- and $B$-mode amplitudes \cite{Collaboration:2014uh}. \HI~orientation preserves the nonunity $EE/BB$ ratio, suggesting that ISM structure is a crucial missing component of these models. The preferential alignment of \textit{Planck} filamentary dust structures with the magnetic field \cite{Collaboration:2015ws} supports this conclusion. Our work underscores the need for a deeper understanding of the interplay between ISM phenomena and polarized dust.

IGW $B$-mode experiments often target the high Galactic latitude sky, where \textit{Planck} data cannot distinguish between the most promising potential targets \cite{Kovetz:2015tc}. Figure \ref{sigmas} shows the relative Galactic latitude dependence of uncertainties in $\theta_{353}$ and $\theta_{RHT}$, where the $\theta_{RHT}$ uncertainty is propagated from the variance in $R\left(\theta\right)$. With sensitive measurements at high latitudes, $\theta_{RHT}$ maps can be used to assess the structure of the magnetic field in targeted regions of sky.

Our results indicate that full foreground templates with higher signal-to-noise than the $Q_{353}$ and $U_{353}$ maps can be constructed by combining $\theta_{RHT}$ with other data describing $P$. A scale-dependent modeling of $p$ and $I$ from a combination of $I_{353}$, $P_{353}$, and \hi data may enable the extension of this work to full polarized dust foreground maps. Such templates should remove CMB and cosmic infrared background emission from $I_{353}$, which we neglect here. We can also replace $I_{353}$ in Eq. \ref{eq.QUtemplate} with an unbiased estimator of $P_{353}$ (e.g. \cite{Plaszczynski:2013gg, 2014arXiv1410.4436V}). $P$ is theoretically determined by the dust column along the line of sight, traced by $I$, and the tangledness of the magnetic field along the line of sight, where more tangled fields cause greater depolarization. $N_{HI}$ is a powerful proxy for $I_{353}$, particularly at high Galactic latitudes where dust emission is low and the expected depletion of \hi into a molecular state is minimal. Changes in $\theta_{RHT}$ for different \hi velocity channels may indicate line-of-sight field tangling, and may elucidate the physical origin of variations in $p$ by isolating components of the magnetic field. This will be the subject of future work, and may lead to further \hi constraints on $P_{353}$. 

In this work we demonstrate that \hi orientation correlates with \textit{Planck} 353 GHz polarization angle. We will process \hi data from the full Arecibo sky in a forthcoming work, as it overlaps with several CMB experiments. Lower resolution \hi surveys such as GASS \cite{McClureGriffiths:2009dn} and EBHIS \cite{Kerp:2011ga} can be used on other regions of the sky, although they do not trace the Galactic magnetic field as precisely as the high resolution GALFA-\hi data \cite{Clark:2014it}. Soon, Galactic all-sky maps from Square Kilometer Array pathfinders \cite{2012MNRAS.426.3385D} will be ideal for \HI-based foreground maps.

\begin{acknowledgments}
\section*{Acknowledgments}
The authors thank David Spergel, Amber Miller, Blake Sherwin, Raphael Flauger, and Kendrick Smith for enlightening discussion of the work presented here. We thank the other members of the GALFA-\hi team for their role in producing that data. We also thank Akito Kusaka and Tobias Marriage for providing survey coordinates on behalf of the PolarBear and CLASS teams, respectively. We thank the anonymous referees for their thoughtful comments. S.E.C. was supported by a National Science Foundation Graduate Research Fellowship under grant No. DGE-11-44155. This work was partially supported by a Junior Fellow award from the Simons Foundation to J.C.H. M.E.P. acknowledges support from NSF grant AST-1410800.
\end{acknowledgments}

\bibliographystyle{apsrev4-1}

\pagebreak
\widetext
\section*{Supplemental Material}

\begin{figure*}[h]
\centering
\includegraphics[trim=4cm 0cm 0cm 0cm, scale=.5]{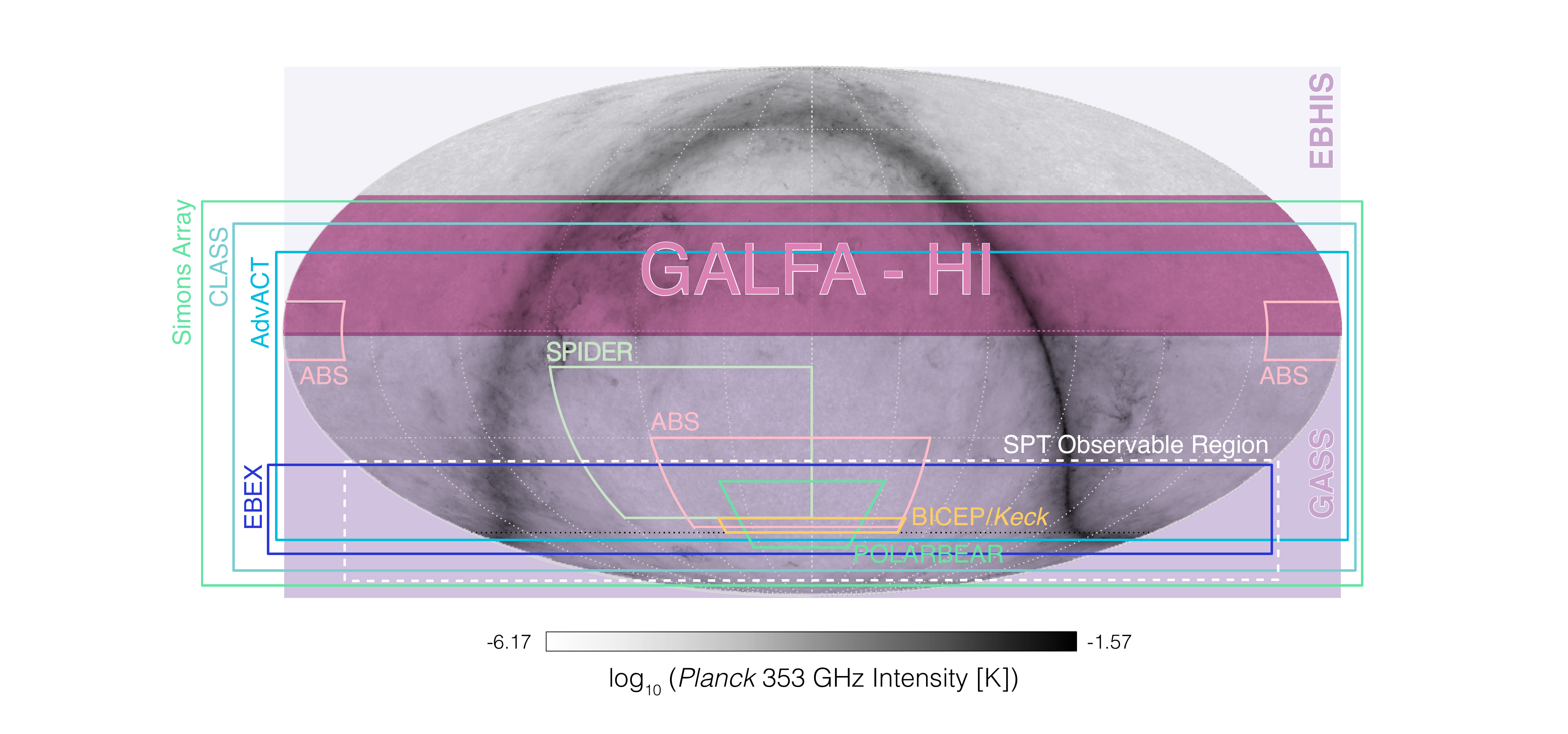}
\caption{Approximate observing regions of various CMB $B$-mode experiments in Equatorial coordinates. Background image is the log of the \textit{Planck} 353 GHz dust intensity [K]. The GALFA-\hi full-sky region is overlaid. The Effelsberg-Bonn \hi Survey (EBHIS) and the Galactic All-Sky Survey (GASS) cover the entire Northern and Southern Equatorial skies, respectively. \label{explocs}}
\end{figure*}

In this work we run the RHT using an unsharp mask kernel diameter $D_K = 15'$, a rolling window size $D_W = 75'$, and an intensity threshold $Z = 70\%$ (see \cite{Clark:2014it}). Under variation of both the velocity channel binning as well as the RHT parameters, our results remain qualitatively unchanged.

Uncertainties for $P$ and $\psi$ are defined by
\beq
\sigma_P = \frac{1}{P} \sqrt{Q^2 {\sigma_{QQ}}^2 + U^2 {\sigma_{UU}}^2}
\eeq
and
\beq
\sigma_\psi = 28.65\degree \sqrt{\frac{Q^2 {\sigma_{UU}}^2 + U^2 {\sigma_{QQ}}^2}{Q^2 {\sigma_{QQ}}^2 + U^2 {\sigma_{UU}}^2}} \cdot \frac{\sigma_P}{P},
\eeq
where we neglect QU covariance. We compute $\sigma_{QQ}^{353}$ and $\sigma_{UU}^{353}$ from half-mission splits of the $Planck$ data, following the procedure outlined in \cite{Collaboration:2015ta}. The analogous quantities $\sigma_{QQ}^{RHT}$ and $\sigma_{UU}^{RHT}$ are computed from the variance in the RHT spectrum, as
\beq
{\sigma_{QQ}^{RHT}}^2 = \int \mathrm{cos}^2 \left(2 \theta \right) \cdot R\left(\theta\right) \, d\theta
\eeq
and
\beq
{\sigma_{UU}^{RHT}}^2 = \int \mathrm{sin}^2 \left(2 \theta \right) \cdot R\left(\theta\right) \, d\theta.
\eeq
We define the difference between $\theta_{353}$ and $\theta_{RHT}$ as
\beq
\delta \theta = \frac{1}{2} \mathrm{arctan} \left[ \frac{\mathrm{sin}(2\theta_{353}) \, \mathrm{cos}(2\theta_{RHT}) - \mathrm{cos}(2\theta_{353}) \, \mathrm{sin}(2 \theta_{RHT})} {\mathrm{cos} (2\theta_{353}) \, \mathrm{cos} (2\theta_{RHT}) + \mathrm{sin} (2\theta_{353}) \, \mathrm{sin} (2 \theta_{RHT})} \right].
\eeq
This equation properly accounts for the $180\degree$ degeneracy in polarization angle.

\end{document}